\documentclass[a4paper,twocolumn,aps,prb]{revtex4}

\usepackage{amsmath}
\usepackage{bbm}
\usepackage{amsfonts}
\usepackage{amssymb}
\usepackage{graphicx}
\usepackage{hyperref}
\usepackage{dsfont}
\usepackage{braket}
\usepackage{color}

\begin{document}

\author{Heng \surname{Wang}}
\author{Guido \surname{Burkard}}
\affiliation{Department of Physics, University of Konstanz, D-78457 Konstanz, Germany}


\title{Mechanically induced two-qubit gates and maximally entangled states for single electron spins in a carbon nanotube}
\begin{abstract}
We theoretically analyze a system where two electrons are trapped separately in two quantum dots on a suspended carbon nanotube (CNT), subject to external ac electric driving. An indirect mechanically-induced coupling of two distant single electron spins is induced by the interaction between the spins and the mechanical motion of the CNT.  We show that a two-qubit iSWAP gate and arbitrary single-qubit gates can be obtained from the intrinsic spin-orbit coupling. Combining the iSWAP gate and single-qubit gates,
maximally entangled states of two spins can be generated in a single step by varying the frequency and the strength of the external electric driving field. The spin-phonon coupling can be turned off by electrostatically shifting the electron wave function on the nanotube. 
\end{abstract}

\pacs{85.85.+j, 71.70.Ej, 76.30.−v, 63.22.Gh}

\maketitle
\section{Introduction}

As  mechanical resonators with potentially high Q factors and large resonance frequencies \cite{Sazonova2004, Huettel2009,Steele2009}, ultra-clean single wall carbon nanotubes (CNTs) are promising systems for studying the coupling of the electron degrees of freedom to the mechanical motion of the resonator\cite{Lassagne2009, Huettel2010, TraversoZiani2011, Ganzhorn2012, Walter2013, Benyamini2014}.
On the other hand the properties of  CNT  such as  valley degeneracy and the curvature induced spin-orbit interaction attract much attention \cite{Ilani2010, Rips2013, E.A.Laird2014}. 
The two valleys in the electron energy spectrum distinguish semiconducting CNT  from   III -V  semiconductors \cite{P'alyi2011}. Qubits can be defined as the electron (hole) spins or the valleys  in  quantum dots (QDs) in CNT  \cite{Palyi2010, Pei2012, Rohling2014}. 
The spin-orbit interaction due to the curvature of CNTs has been studied both in theory \cite{Huertas-Hernando2006, Jeong2009, Izumida2009} and observed in the laboratory\cite{Kuemmeth2008,Steele2013}. 
The spin-orbit interaction plays an important role as a source of spin decoherence \cite{Bulaev2008}  and at the same time it allows the electrical control of the spin in bent CNT   in a magnetic field as well as cooling of the CNT resonator using spin-polarized current \cite{Flensberg2010,Stadler2014}. 
Furthermore, the coupling of the spin in a single QD and the deflection of the CNT was studied \cite{Rudner2010}, and the spin-phonon coupling, which is induced from the spin-orbit coupling where the tangent vector instantaneously depends on the phonon displacement, provides a new platform for operating spins and quantized flexural modes \cite{P'alyi2012}. 
The read-out of the resonator vibration frequency and the detection of the single electron spin in the QD have been proposed based on the spin-phonon coupling  \cite{Ohm2012a, Struck2014}.

\begin{figure}[t]
	\begin{center}
	 {\includegraphics[width=0.48\textwidth]{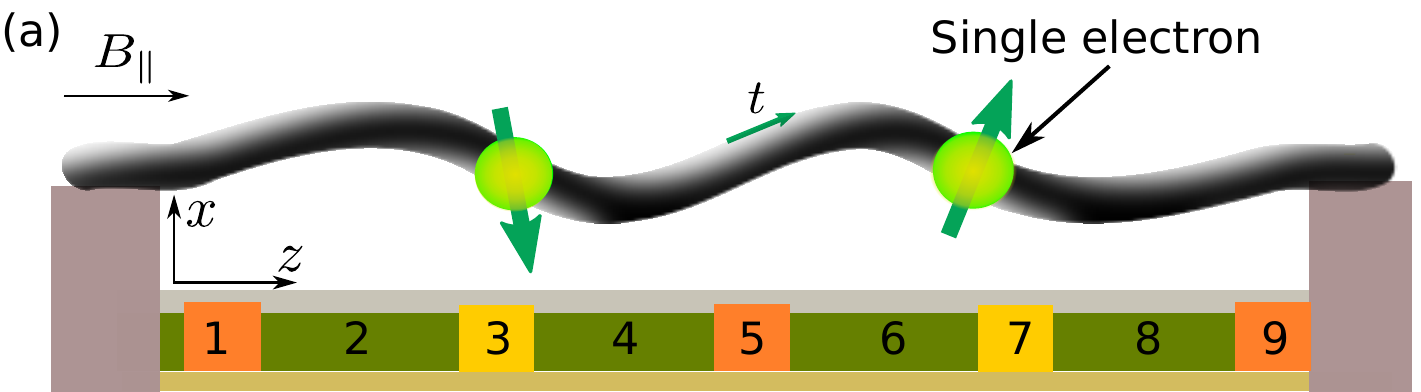}} 
 {\includegraphics[width=0.48\textwidth]{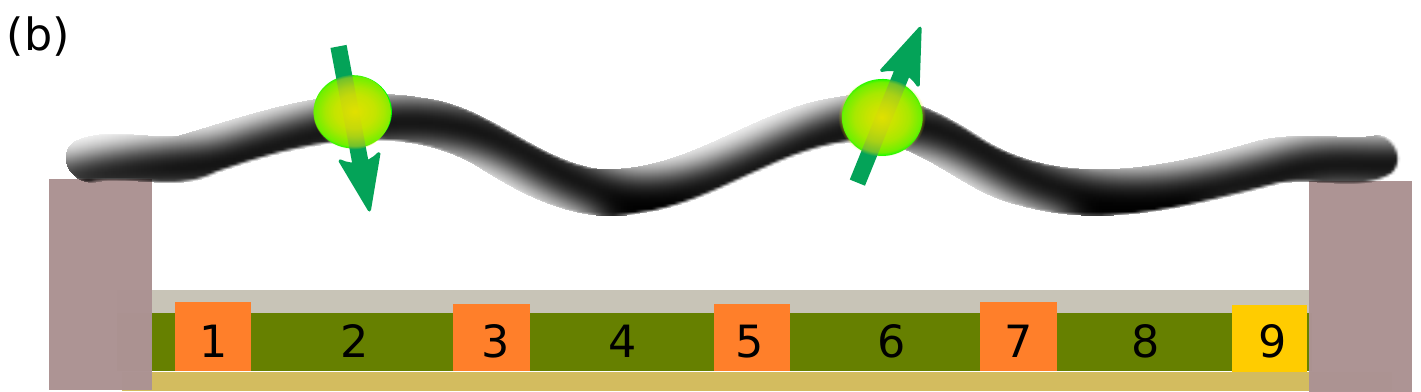}}
\end{center}
\caption {(a) Schematic of the nanomechanical system where two quantum dots (QDs) lie in a doubly clamped, suspended CNT which is fixed by two supports at two ends. QDs are formed by the  electronic potentials applied by the gate electrodes no. $1$, $5$ and $9$   to form two QDs. Here, we assume that the fourth harmonic flexural mode of the CNT is excited. The standing wave in each QD is asymmetric  and has one single electron trapped inside. The charged CNT is driven to vibrate along the $x$ axis by an external ac electric field applied by an antenna or the gate electrodes.  A magnetic field is applied along the $z$ axis. 
(b) To switch off the coupling between two QDs, the QDs can be electrostatically shifted. The left QD between gates no. $1$ and no. $3$ as well as the right QD between gates no. $5$ and no. $7$ are both left-right symmetric.    }
\label{fig:switch}
\end{figure}

Previously, we have   proposed arbitrary   single  qubit gates  using an electron spin in a single QD which lies in a suspended CNT making use of the spin-phonon coupling of  the mechanical motion of the CNT \cite{WangandBurkard2014}. 
For quantum information and quantum computation, one-qubit and two-qubit gates are universal \cite{DiVincenzo1995}. Individual two-qubit gates that can form a universal set in combination with single-qubit gates are e.g. CNOT, $\sqrt{{\rm SWAP}}$ and iSWAP\cite{DiVincenzo13101995,Loss1998, Schuch2003}. %
 There are approaches related to inhomogeneous magnetic fields to produce universal gates of spins and to achieve coupling of long distance spins in NV centers using mechanical resonators \cite{Xu2009,Zhou2010,Chotorlishvili2013}. Compared with the use of inhomogeneous magnetic fields, electric fields are easier to control temporally and spatially.  Universal quantum computation requires that arbitrary pairs of two qubits can interact with each other. It is usually not easy to fulfill this requirement because long distance coupling can be very demanding. 
In the present paper we theoretically study a two-qubit iSWAP gate and arbitrary single qubit gates in a nano-mechanical scheme where two electrons are trapped separately in two QDs on a suspended CNT.  The indirect coupling of two distant single-electron spins in two separated dots is mediated by the vibrational motion of the CNT.  A single-step preparation of maximally entangled states is obtained by combining the iSWAP gate and single-qubit gates. All quantum gates proposed here can be controlled electrically. We show that the spin-phonon coupling in each QD can be turned off by electrostatically shifting the electron wave function on the nanotube. We predict that arbitrary pairs of QDs can be coupled through the spin-phonon coupling on CNT with multiple QDs. When there are more than two QDs in the CNT, it is possible to couple arbitrary pairs of distant electron spins and at the same time turn off the spin-phonon coupling in the other QDs.

This paper is organized as follows. In Sec.~\ref{sec:model} the nanomechanical system and the model Hamiltonian are introduced. In Sec.~\ref{subsec:schrieffer-wolff transformation} a Schrieffer-Wolff transformation is applied to obtain the effective Hamiltonian in spin space and  to obtain the iSWAP gate from the effective Hamiltonian. In Sec.~\ref{sec:iSWAP and Entangled states},  we determine the wave function of the qubit state by solving  the Schr\"odinger equation and simulate numerically the fidelity of the maximally entangled states in a open quantum system by using  a quantum master equation. In the Sec.~\ref{sec:switch} we describe how to shift the electron wave function to turn off the spin-phonon coupling.


\section{Model}
\label{sec:model}

We assume that a doubly clamped, suspended CNT is fixed  on two supports at both ends \cite{Steele2009,Lassagne2009}, see Fig. \ref{fig:switch}   (a).   Two QDs can be formed by applying proper voltages on gate electrodes below the suspended CNT. We further assume that a voltage  is applied to the gates  no. $1$, $5$ and $9$ so that the electrons are prevented from tunneling out of the CNT or from tunneling between QDs. The other gates  can be used to tune the resonance frequency of the CNT. We assume that two electrons are trapped in two QDs \cite{Benyamini2014}, separately, and that an external longitudinal magnetic field $\mathbf{B}_{\parallel}$ is applied along the $z$-axis of the CNT. An external ac electric field is applied by an antenna on the top or on the gates  to excite the vibration of the charged CNT.


Here,   two single electron spins in two QDs are assumed to couple to the the vibrational motion simultaneously, hence these two spins are indirectly  coupled via phonon exchange.
We describe this system using the Hamiltonian  
\begin{eqnarray}
H &=& H_0 +H_1 ,\label{eq:totalHamiltonian}\\
H_0 &=& \sum_i \frac{\hbar \omega
_{{\rm q}i}}{2}\sigma_{zi}+\hbar \omega_{\rm p}a^{\dag}a,\\
H_1 &=&  2 \hbar \lambda(a+a^{\dag}) \cos(\omega t)+\sum_i\hbar g_i (a+a^{\dag})(\sigma_{+i}+\sigma_{-i}),\nonumber\\
\end{eqnarray}
where $i= 1, 2$ refers to two electrons in two separate QDs. The two spin states cross at the magnetic field $B^*\approx \Delta_{ \rm so}/(2 \mu_B)$ in the $K$ valley of the ground state of a single electron QD in CNT\cite{P'alyi2012, Ohm2012a, Struck2014, WangandBurkard2014}, where $\Delta_{so}$ is the spin-orbit coupling strength and $\mu_B$ is the spin magnetic moment.
 We choose these two spin states as the qubit and assume that we are near the crossing point. The Zeeman splitting energy  between qubits  induced by the magnetic field $\mathbf{B}_i$ is  $\hbar\omega_{qi}=g\mu_B( B _i-  B ^*)$.
The quantized mechanical motion is described by the phonon mode with frequency $\omega_{\rm p}$ and $a$ ($a^{\dagger}$) is the phonon annihilation (creation) operator. We assume the system to be at low temperature $\hbar \omega_{\rm p} \leq k_{\rm B}T$.  Here $\sigma_{zi}$ is the Pauli $z$ matrix of the electron spins and $\sigma_{\pm i}$ are the corresponding spin raising and lowering operators. For simplicity, we only consider the fourth excited quantum harmonic flexural mode along the $x$ axis of the CNT in the present paper.

Due to the curvature caused by the vibrational motion, the local tangent vector $\mathbf{t}$ of the CNT   depends on the displacement coordinate\cite{Rudner2010}, and it induces an interaction between the mechanical motion and the electron spin.
Spin-phonon coupling   originates from the dynamical spin-orbit interaction  $\boldsymbol{\sigma} \cdot \mathbf{t }(z)=\sigma_{z}+({\rm d}u/{\rm d}z)\sigma_x$, where $u(z) $ is the displacement at the coordinate point $z$, $u(z)=f(z)\frac{l_0}{\sqrt{2}}(a+a^{\dagger})$, where $f(z)$ is the waveform of the phonon mode and $l_0$ is the zero-point displacement. For different QDs in the nanotube,  the spin-phonon coupling strengths are $g_{i}=\Delta_{so} \braket{f'(z)}_i l_0/2\sqrt{2}$. Here $f'(z)$ is the derivative of the waveform of the phonon mode.  Considering the electron distribution, we obtain the average of the derivative of the waveform $\braket{f'(z)}_i=\int^{l_i/2}_{-l_i/2} {\rm d}z \frac{{\rm d} f(z)}{{\rm d}z}D_{i}(z)$ where $D_i(z)$ are the charge densities of two QDs and each quantum dot is between $-l_i/2$ and  $l_i/2$. We use realistic parameter $L=800 \ {\rm nm}$, $\Delta_{so}=370 \ {\rm \mu eV}$, $l_0=2.5 \ {\rm pm}$ and obtain the 
value of the coupling strength $g/(2\pi)=0.56 \ {\rm MHz}$ 
\cite{P'alyi2012}.

We assume that $D_i(z)$ are symmetric functions in the QDs.  In this case, 
the spin-phonon coupling strength is non-zero when the parity of $ f(z)$ is odd in 
the QD.
In other words, if the parity of the charge density function is even in a QD, to avoid canceling out the spin-orbit interaction, there should be a left-right asymmetric standing wave in the QD \cite{P'alyi2012}.

  \begin{figure}[t]	
                   {\includegraphics[width=0.45\textwidth]{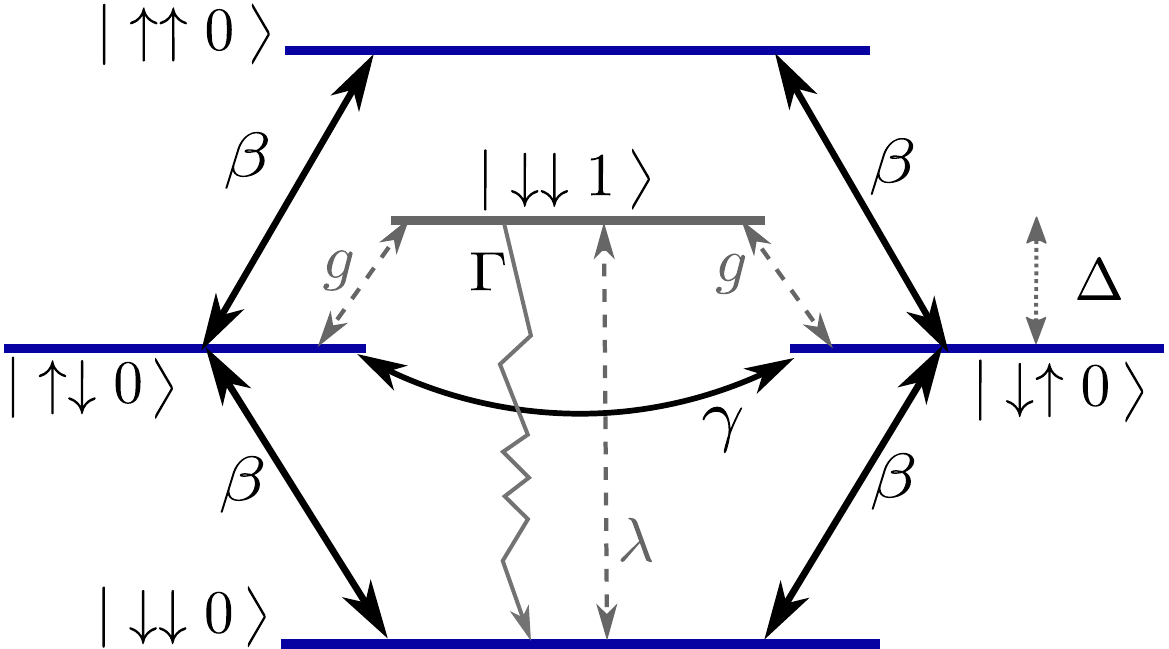}}                 
\caption {The energy-level diagram of the combined two-qubit and phonon system.  $\ket{\sigma \sigma ' n}$ denotes a state, where $\sigma \ (\sigma ')$ represents the first (second) spin state and $n$ is the number of phonons. The dashed lines denote the coupling strength of the external ac electric field $\lambda$ and the spin-phonon coupling strength  $g$. The ac electric field is detuned from the phonon frequency by $\Delta$. The coherent coupling between states $\ket{\downarrow\downarrow 0}$ and $\ket{\uparrow\downarrow 0}$ is mediated by the state $\ket{\downarrow\downarrow 1}$ through the driving and the spin-phonon coupling. The effective single spin resonance coupling  strength $\beta$ and the effective two-spin interaction strength $\gamma$   are obtained by   deriving the effective Hamiltonian for the $ n=0$ subspace and thereby eliminating the $\ket{\sigma\sigma '1}$ states with a Schrieffer-Wolff transformation (similarly for $ n>0 $). Here $\Gamma$ is the damping rate of the CNT.   }  
  \label{fig:energylevel}
\end{figure}

\section{Effective Hamiltonian from Schrieffer-Wolff transformation}
\label{subsec:schrieffer-wolff transformation}

For a better understanding of the evolution of the spins, we derive the effective Hamiltonian in the subspace of  the spins.  We assume that the difference between the phonon energy and the qubit energy is much larger than the spin-phonon coupling strength and the driving strength, i.e., that $\omega_{\rm p}-\omega_{{\rm q}i}\gg g_i,\lambda$. The Schrieffer-Wolff transformation can be applied when the subspaces with different phonon numbers  are energetically well separated. First, we obtain the effective Hamiltonian in the lowest subspace with zero phonon, then we can use the same method to obtain the effective Hamiltonian in all subspaces. 
  The effective Hamiltonian from the time-dependent Schrieffer-Wolff transformation in the lowest phonon subspace can be written as \cite{Goldin2000,WangandBurkard2014}
\begin{eqnarray}
H_{\rm  eff}=&H_{\rm  eff}^0+H_{\rm  eff}^1, \label{eq:SchriefferSchrodinger}\\
H_{\rm eff}^0 =&
\sum_i \zeta_i \sigma_{zi}
+\gamma \sigma_{x1}\sigma_{x2},\\
H_{\rm eff}^1 =& 2 \cos{\omega t} \sum_i ~ \beta_i  \sigma_{xi},
\end{eqnarray}
where 
\begin{eqnarray}
\zeta_i &=&\frac{ \hbar\omega_{{\rm q}i}}{2}-\frac{(2n+1) \hbar\omega_{{\rm q}i} g_i^2 }{\omega^2_{\rm p}-\omega_{{\rm q}i}^2},\\
\beta_i &=& -\frac{\hbar\lambda g_i\omega_{\rm p}(\omega^2-2\omega_{\rm p}^2+\omega_{{\rm q}i}^2 )}{(\omega^2-\omega_{\rm p}^2)(\omega_{\rm p}^2-\omega_{{\rm q}i}^2)},\\
\gamma &=& -\frac{\hbar g_1 g_2 \omega_{\rm p}(-2 \omega_{\rm p}^2+ \omega_{\rm q1}^2+ \omega_{\rm q2}^2)}{(\omega_{\rm p}^2-\omega_{\rm q1}^2)(\omega_{\rm p}^2-\omega_{\rm q2}^2)}.
\end{eqnarray}
It is worth pointing out that in Eq.(\ref{eq:SchriefferSchrodinger}) there is not only the coupling term which denotes  coupling of two spins, but also the single electron spin rotation terms $\sigma_x$  and  $\sigma_z$.

\begin{figure}[t]	
\begin{center}
 {\includegraphics[width=0.45 \textwidth]{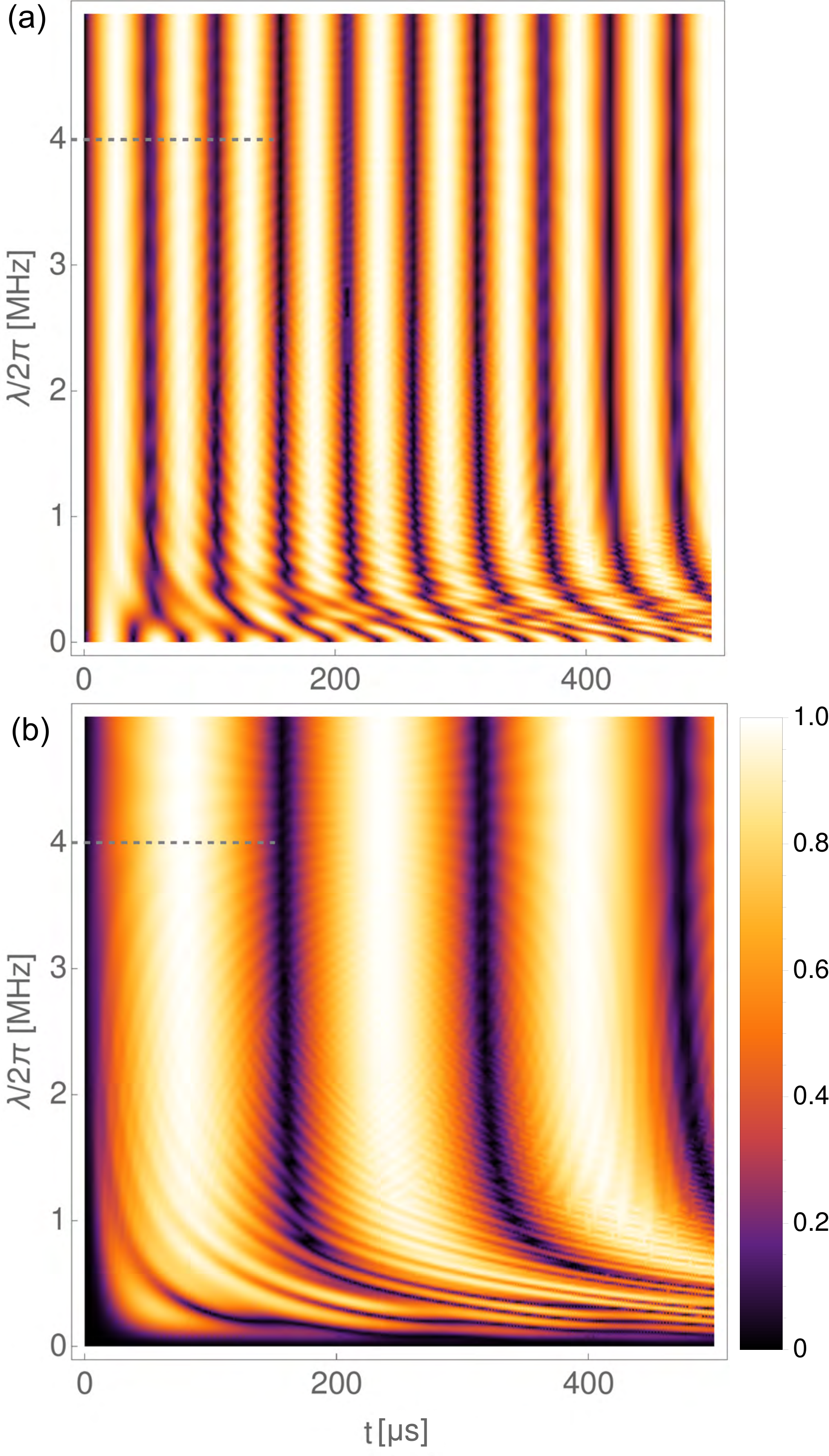}}
                \end{center}
\caption {The time evolution of the concurrence (color scale) as a function of the driving strength $\lambda$ with different initial states (a) $\ket{\uparrow\downarrow}$  and (b)  $\ket{\uparrow\uparrow}$.  The dashed lines denote the time evolutions of the concurrence with a fixed driving strength which are shown in Fig. \ref{fig:wavefunction1}.  When  $ \lambda=0$, in (a) the period between high concurrence peaks is $t_0 =\pi (\omega_{\rm p}^2-\omega_{\rm q}^2)/ (4g^2\omega_{\rm p})$,  which depends on the strength $g$ of the spin phonon coupling, while in (b), the spin-phonon coupling does not have any effect on initial state  $\ket{\uparrow\uparrow}$. When $\lambda \gg g$, the period of the concurrence peaks depends on  the initial state  and the periods are $t\approx   (4/3)t_0$ in (a)
and $t \approx   4 t_0$ in (b). 
  The system parameters are chosen to be $ \omega_{\rm p}/ (2\pi)= 1500 \ {\rm MHz}, \ \omega / (2\pi)=  \omega_{\rm q}/ (2\pi)=  1450 \ {\rm MHz}, \ g /(2\pi) =  0.56\ {\rm MHz}$ and $ n=0$. }
  \label{fig:concurrence}
\end{figure}

To get rid of the time dependence in $H_{\rm eff}^1$, we transform the Eq.(\ref{eq:SchriefferSchrodinger}) into the rotating frame with frequency $\omega$, using the transformation $H_{\rm eff}^I=UH_{\rm eff}U^\dagger -iU \dot{U}^{\dagger}$ with $U=e^{i (\omega/2)t  \sum_i\sigma_{zi} }$. We assume $ \omega_{\rm p}\sim \omega_{\rm q} \sim \omega$, $\Delta=\omega_{\rm p}-\omega$ and $\Delta \gg g$.
 The fast oscillating terms with  $e^{\pm 2i\omega  t} $ can be dropped in the rotating-wave approximation.   We extend our analysis to the full phonon space and obtain the effective Hamiltonian from the Schrieffer-Wolff transformation in the rotating frame  (see Appendix \ref{Appsec:Schrieffer}) 
\begin{equation}
H'_{\rm eff}=\sum_i (\alpha_i\sigma_{zi}+\beta_i\sigma_{xi}) +\gamma( \sigma_{+1}\sigma_{-2}+ \sigma_{-1}\sigma_{+2}),
\label{eq:SchriefferInteraction}
\end{equation}
where 
\begin{equation}
\alpha_i=\frac{ \hbar\omega_{{\rm q}i}}{2}-\frac{(2n+1) \hbar\omega_{{\rm q}i} g_i^2 }{\omega^2_{\rm p}-\omega_{{\rm q}i}^2}-\frac{\hbar\omega}{2},
\end{equation}
 and $n=a^{\dagger}a$ is the phonon number operator.    
The energy-level spectrum is shown in Fig. \ref{fig:energylevel}. 
 Arbitrary single-qubit gates of the single electron spin can be obtained   by combining rotations about the $x$-axis and   the $z$-axis \cite{WangandBurkard2014}. The rotations about the $z$-axis of each QD can be adjusted by changing the driving frequency, and  can be switched off by setting $\omega=\omega_{\rm qi}(1-\frac{2(2 n +1)g_i^2}{  \omega^2_{\rm p}+\omega^2_{\rm qi}})$.  The rotations about the $x$-axis can be adjusted by choosing different strengths of the driving field, and it can be switched off by setting $\lambda=0$.   From rotations about $x$-axis ($z$-axis), $X$ ($Z$) gate which is $X=\sigma_x$ ($Z=\sigma_z$) can be obtained. \cite{WangandBurkard2014}.

We show how to obtain the iSWAP gate from the effective Hamiltonian. We assume $\omega_{\rm i}=\omega_{\rm q}$, $\omega=\omega_{\rm q}$ and $g_i=g$. 
The iSWAP gate is obtained in the absence of driving, $\lambda=0$ where the phonon vacuum fluctuations couple the two QDs. Choosing the appropriate pulse length $ t = \pi (\omega_{\rm p}^2-\omega_{\rm q}^2)/ (4g^2\omega_{\rm p})$,  we obtain the evolution operator $U'=e^{-iH'_{\rm eff}  t/\hbar}$ in the basis   $\{\ket{\uparrow \uparrow }, \ket{\uparrow \downarrow },\ket{\downarrow \uparrow }, \ket{\downarrow \downarrow }\}$ in the following form \cite{Zhou2010} 
\begin{equation} 
{U'}=\left( 
\begin{array}{cccc}
e^{-i  \frac{\pi\omega_{\rm q}}{2 \omega_{\rm p}}   } & 0 & 0& 0  \\
0&0& i&0\\
 0&i& 0&0\\
0 &0& 0& e^{ i  \frac{\pi\omega_{\rm q}}{2 \omega_{\rm p}} }
\end{array} \right).
\label{eq:smatrix}
\end{equation} 
We can see from Eq.(\ref{eq:smatrix}) that the evolution operator is an iSWAP gate with relative phases.  We can apply  single qubit gates $\sigma_{z i}$ on the two QDs for $t=\pi(2/\omega_{\rm q}-1/2\omega_{\rm p})$ to eliminate the relative phases between states $\ket{\uparrow\uparrow}$ and $\ket{\downarrow\downarrow}$.

\begin{figure}[t]	
                       {\includegraphics[width=0.47\textwidth]{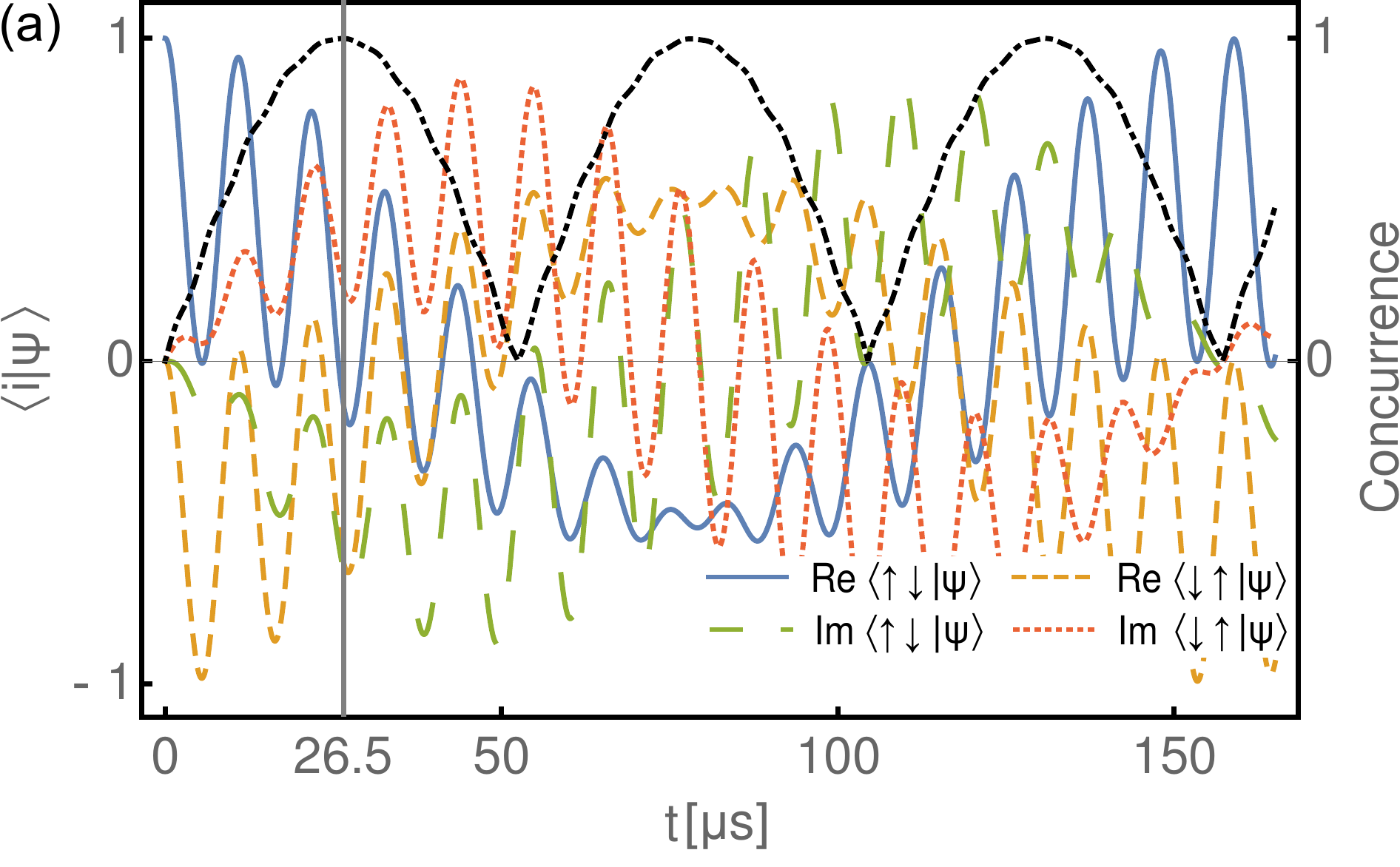}} 
                       {\includegraphics[width=0.47\textwidth]{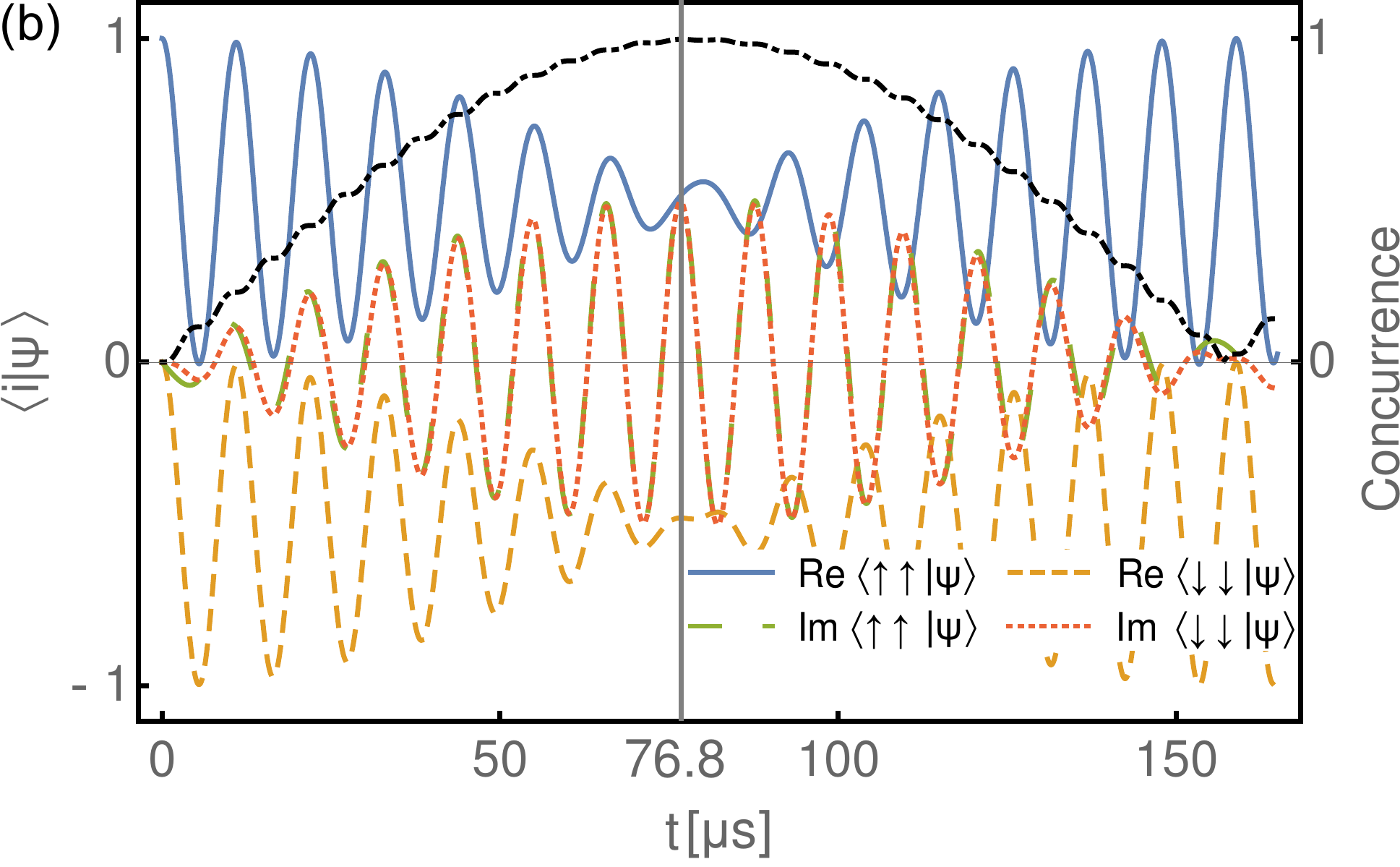}}
\caption {The time evolution of the coefficients $c_i(t)=\braket{i|\psi(t)}$ of the wave function  and the concurrence with driving strength $\lambda/(2\pi)= 4\ {\rm MHz}$ for different initial states, (a) $\ket{\uparrow \downarrow }$,  and (b) $\ket {\uparrow \uparrow} $. (a) The maximally entangled state obtained at $t_1=26.5 \ { \rm \mu s}$  is  $\psi(t_1)\approx(-0.09+0.23i)\ket{\uparrow \uparrow}+(-0.13-0.65i)\ket{\uparrow \downarrow}+(-0.62+0.23i)\ket{\downarrow \uparrow}+(-0.16+0.21i)\ket{\downarrow \downarrow}$.  (b) The maximally entangled state obtained at $t_2=76.8 \ { \rm \mu s}$  is $\psi(t_2)\approx(0.52+0.51i)\ket{\uparrow \uparrow}+(-0.02i)\ket{\uparrow \downarrow}+(-0.02i)\ket{\downarrow \uparrow}+(-0.48+0.49i)\ket{\downarrow \downarrow}$. The other parameters are the same as in Fig.\ref{fig:concurrence}. }
  \label{fig:wavefunction1}
\end{figure}

\section{Wave function and maximally entangled states}
\label{sec:iSWAP and Entangled states}

One can combine the iSWAP and single qubit gates to obtain any maximally entangled states from any initial product state. In our case, we can achieve a single-step preparation of maximally entangled state  by adjusting the driving strength and driving frequency. The effective Hamiltonian in Eq.(\ref{eq:SchriefferInteraction}) is time independent.  We solve the Schr\"odinger equation  $ i\frac{\partial (\ket{n}\otimes\ket{\psi(t)})}{\partial t}=H'_{\rm eff}(\ket{n}\otimes\ket{\psi (t)})$  in the subspace with phonon number $n=0$, where $\ket{\psi(t)}=c_1(t)\ket{\uparrow\uparrow}+c_2(t)\ket{\uparrow\downarrow}+c_3(t)\ket{\downarrow\uparrow}+c_4(t)\ket{\downarrow\downarrow}$ is the wave function of the qubit states with a initial product state. The exact maximally entangled states and the corresponding time points can  both be obtained by solving the Schr\"odinger equation. 
The Hamiltonian in the product basis is 
\begin{equation}
\begin{split}
H'_{\rm eff}=\left(\begin{array}{cccc}
2\alpha &\beta & \beta& 0\\
\beta &0 & \gamma & \beta\\
\beta &\gamma & 0 & \beta\\
0 &\beta & \beta& -2\alpha
\end{array} \right).
\end{split}
\end{equation}
We can use the four eigenvalues $\mu_i(\Delta, \lambda, g)$ and four eigenstates $\ket{\psi_i}$ to obtain the general solution of the wave function  $\ket{\psi(t))}=A_1\ket{\psi_1(t))} e^{-i\mu_1 t}+A_2\ket{\psi_2(t))} e^{-i\mu_2 t}+A_3\ket{\psi_3(t))} e^{-i\mu_3 t}+A_4\ket{\psi_4(t))} e^{-i\mu_4 t}$, where $A_i$ is dependent on the initial state. 
 
To quantify the generated entanglement, the concurrence, as a measure of entanglement, is evaluated. The concurrence for pure two-qubit state can be written as $C(\ket{\psi(t))}=|\braket{\psi(t)|\sigma_y \otimes \sigma_y|\psi(t)^*}|=2|c_1(t)c_4(t)-c_2(t)c_3(t)|$.
 We show the concurrence as a function of time and driving strength for different initial states in Fig. \ref{fig:concurrence}. Since the concurrence reaches the value $C=1$, we know that maximally entangled states can be obtained. From the solution of the Schr\"odinger equation,   we obtain the time dependent $C(t)$ of the concurrence. The period of the concurrence depends on the initial state, the driving strength and the spin-phonon coupling strength.  Varying in time the driving strength, we can shift the maximum of the concurrence. In the following we take two initial states as examples in   Fig. \ref{fig:concurrence}. When  $\lambda=0$, the period of the concurrence is  $ t = \pi (\omega_{\rm p}^2-\omega_{\rm q}^2)/ (4g^2\omega_{\rm p})$  and only the iSWAP gate acts on the initial state.  The iSWAP gate creates entanglement on the initial state $\ket{\uparrow\downarrow}$ in Fig. \ref{fig:concurrence} (a) but no effect with initial state $\ket{\uparrow\uparrow}$ in Fig. \ref{fig:concurrence} (b). When the driving 
strength is nonzero, the electron spin resonances   
corresponding to $X$-
gates are  on.  When $0<\lambda \leq g$, the frequency of the electron spin resonances is  slower than that of the iSWAP gate, therefore the period between high concurrence peaks depends on the spin resonance frequency. We can see from Fig. \ref{fig:concurrence}  
that the frequencies of the blurred oblique lines 
in Fig. \ref{fig:concurrence} (a) 
and the bright oblique lines in Fig. \ref{fig:concurrence} (b) are determined by the strength 
$\beta $ of the single electron rotation $\sigma_x$. 
When  $ \lambda > g$, the frequency of the spin resonances is higher than  for the iSWAP gate. The electron spin resonances contribute a fast oscillation of the wave function. The periods of the concurrence depend on the strength of the iSWAP gate.  Corresponding to the vertical strips in Fig. \ref{fig:concurrence},  the periods of the concurrence cycles are $t \approx \pi (\omega_{\rm p}^2-\omega_{\rm q}^2)/ (3g^2\omega_{\rm p})$ in Fig. \ref{fig:concurrence} (a) and  $t \approx \pi (\omega_{\rm p}^2-\omega_{\rm q}^2)/ ( g^2\omega_{\rm p})$ in Fig. \ref{fig:concurrence} (b). 
We also show the time evolution of the wave function with driving strength $\lambda/(2\pi)=4\ {\rm MHz}$ in Fig. \ref{fig:wavefunction1}, which corresponds to the dashed line in  Fig. \ref{fig:concurrence} with $ \lambda \geq g$.  Although the coefficients of the wave function are fast oscillating due to the strong rotation $\sigma_x$,  the envelopes of the time evolution of coefficients still correspond to the iSWAP gate. 
  \begin{figure}[t]	
                   {\includegraphics[width=0.47\textwidth]{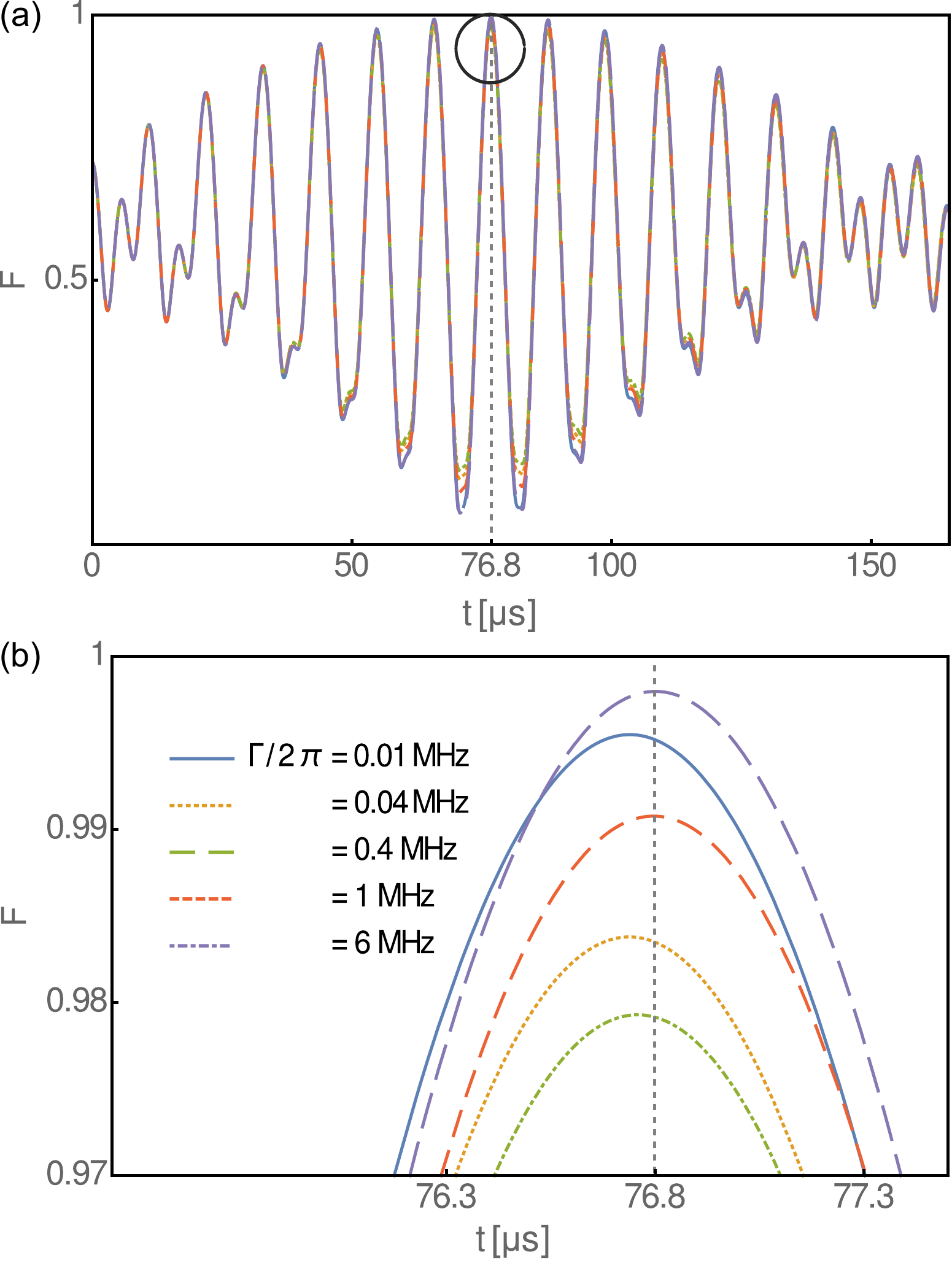}} 
\caption { (a) The time evolution of the fidelity  $F =\sqrt{ \braket{\Psi|\rho_s|\Psi} }$ with driving strength $\lambda/(2\pi)=4.0 \ {\rm MHz}$ with a maximum near $ t_{\rm ideal}=76.8 \ { \rm \mu s}$ obtained by solving the master equation Eq.(\ref{eq:master_eqn_finiteT}) which takes into account the thermal equilibrium phonon bath at $T=30\ {\rm mK}$ and the damping rate of the resonator. We magnify the circle in (b). (b) The fidelity for the case without damping is smaller than $1$ due to the finite temperature. The shift between the peaks and the dashed line decreases when the damping rates increase. 
We truncate the phonon Hilbert space for $n>6$.  The other parameters are the same as in Fig. \ref{fig:concurrence}. }
  \label{fig:Fidelity}
\end{figure}

To include the damping of the CNT due to the coupling to a thermal bath at temperature $T$, we use a master equation for the nonunitrary dynamical simulation.
The density of states of the other phonon modes  is small, therefore we can expect a  small spontaneous qubit relaxation rate $1/T_1$ and neglect it in the following.  We obtain the master equation for the density matrix $\rho$,
 \begin{equation}
	\begin{split}
	\dot{\rho}=& -\frac{i}{\hbar}[H,\rho] +(n_{B}+1)\Gamma \left(a\rho a^{\dagger} -\frac{1}{2}\{a^{\dagger} a,\rho\}\right) \\
	& +n_{B}\Gamma\left(a^{\dagger}\rho a-\frac{1}{2}\{a a^{\dagger},\rho\}\right),
        \end{split}
	\label{eq:master_eqn_finiteT}
\end{equation}
where $n_{B}=1/(e^{\hbar \omega_{\rm p}/k_{\rm B}T}-1)$ is the Bose-Einstein occupation factor,  and  $\Gamma \ll g$ is the damping rate of the CNT. The phonons follow the Bose-Einstein statistics in the thermal equilibrium in the initial state of the density matrix that $ \rho=\frac{1}{Z}\sum_{n=0}^{ \infty} e^{-n\hbar \omega_{\rm p}/(k_{\rm B}T)}\ket{n}\bra{n}\otimes\ket{\psi }\bra{\psi }$, where $Z=\sum_{n=0}^{ \infty} e^{-n\hbar \omega_{\rm p}/(k_{\rm B}T)}$ is the partition function.
The total spin state is given by the partial trace over the phonons $\rho_{\rm s}={\rm Tr}_{\rm ph}\rho$.  The fidelity relative to the two-qubit entangled state is defined as $F =\sqrt{\bra{\Psi} \rho_{\rm s}\ket{\Psi}}$. 

  \begin{figure}[t]	
                   {\includegraphics[width=0.47\textwidth]{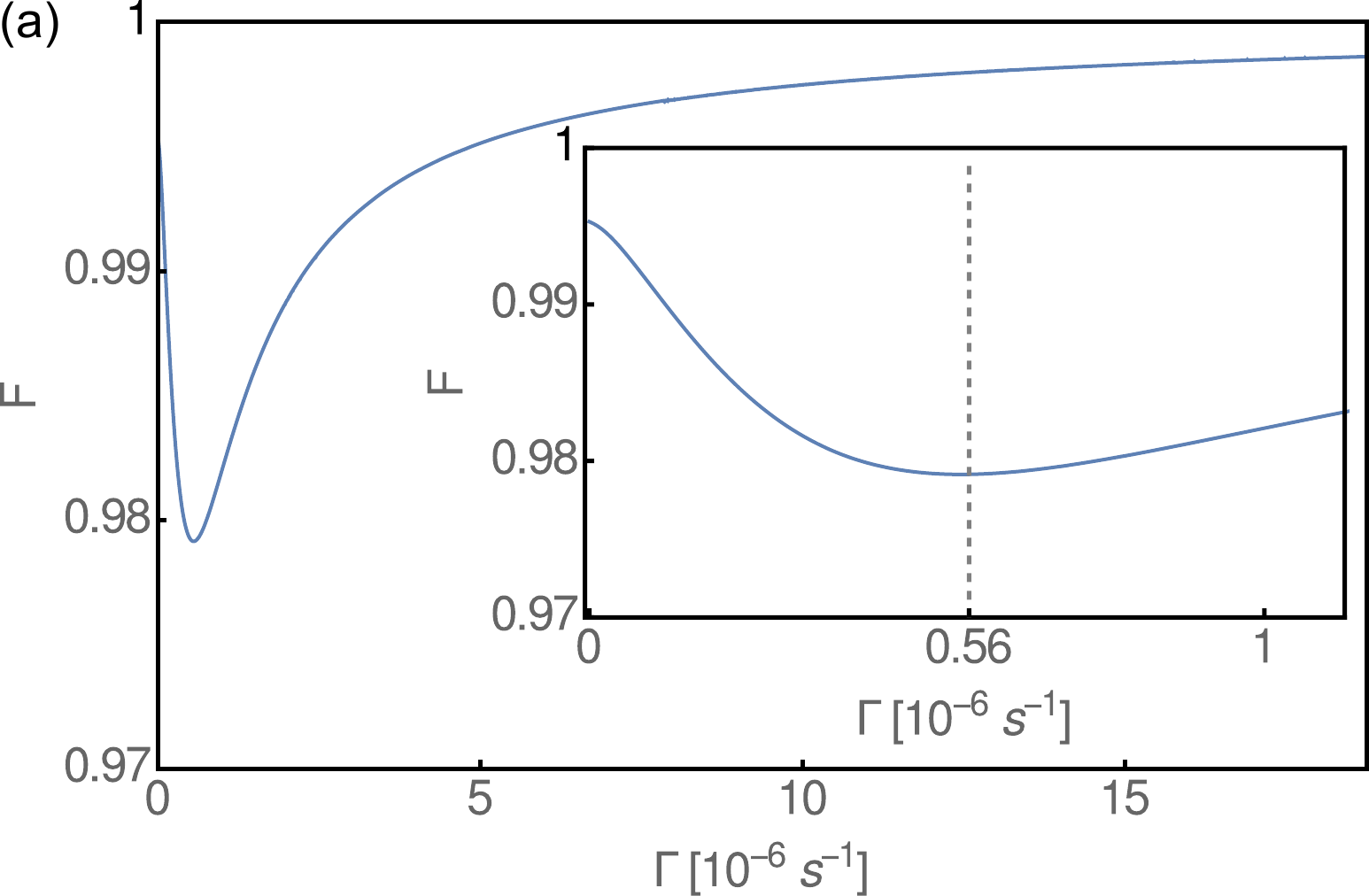}}
                   {\includegraphics[width=0.47\textwidth]{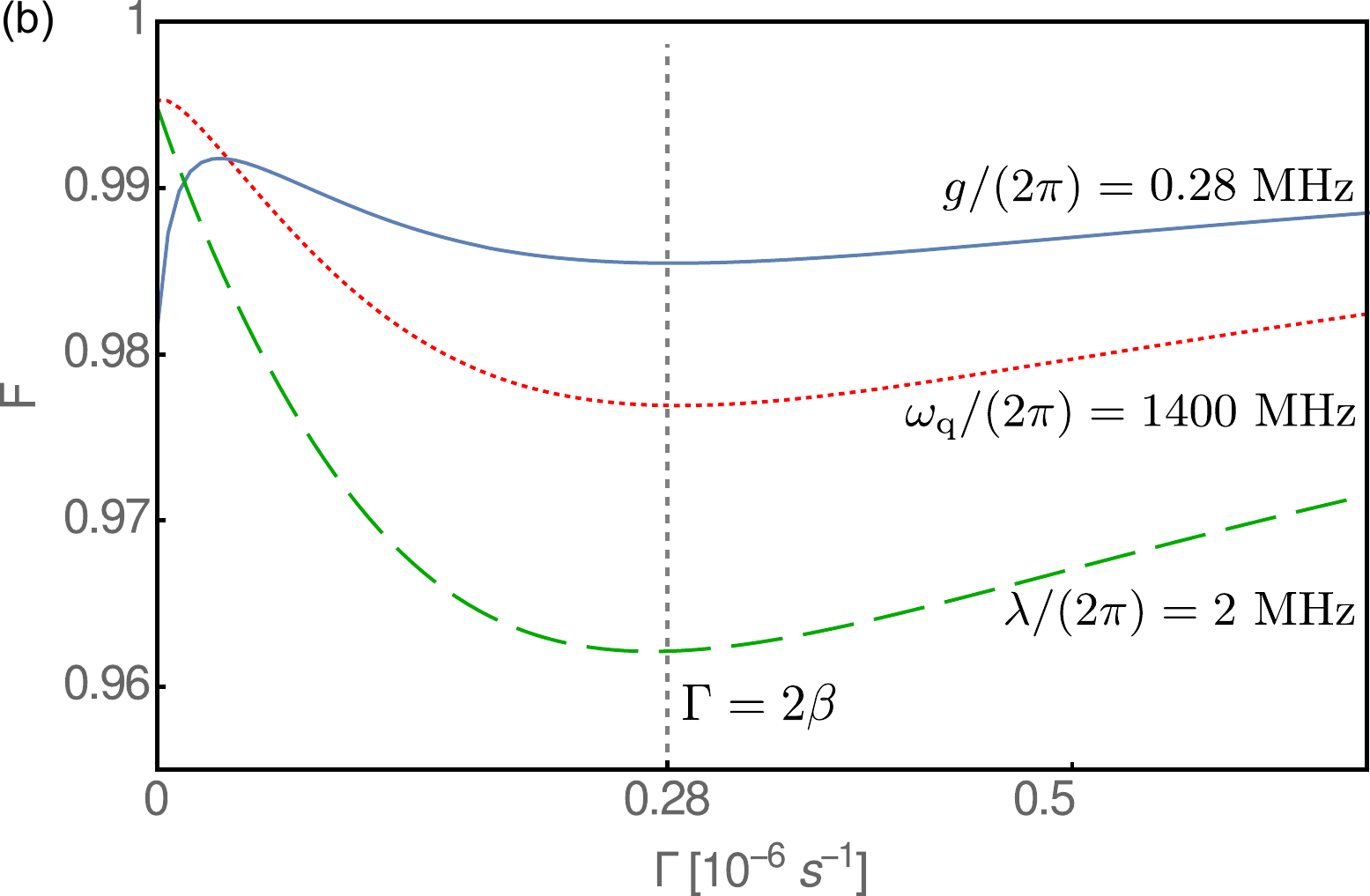}}                 
\caption { (a)  The fidelity $F$ of obtaining the maximally entangled state $\ket{\Psi}$ as a function of the damping rate $\Gamma$ with initial state $\ket{\uparrow\uparrow}$, taking into account the phonon bath in the thermal equilibrium at $T=30 \ {\rm mK}$. The initial state is $\ket{\uparrow\uparrow}$.   The stochastic resonance occurs  at $\Gamma=2\beta=0.56  \times {10^{-6}} \ {\rm s^{-1}}$.
 The other parameters are the same as in Fig. \ref{fig:concurrence}.  The fidelity for $\Gamma=0$ is limited by the finite temperature $T>0$. (b)  The fidelity $F$ of obtaining the maximally entangled states with different parameters as a function of the damping rate $\Gamma$. We  separately modify the parameters $\lambda$, $\Delta $ and $g$ by a factor of 2, as $\lambda/(2\pi)=2 \ {\rm MHz}$,   $g/2\pi=0.28 \ {\rm MHz}$,  or $ \omega/ 2\pi=\omega_q/ 2\pi=1400 \ {\rm MHz}$, while keeping other parameters as in (a). Hence the value of $\beta\approx \lambda g / \Delta$ is the same for all three cases, and half of its value in (a).  The maximally entangled states in these three cases are different and obtained at a different time. The minimal fidelities in all these cases occurs at $\Gamma=0.28  \times {10^{-6}} \ {\rm s^{-1}}$, which shows that the maximum of the stochastic resonance is at $\Gamma \approx 2\beta  $ (vertical dashed line). We truncate the phonon Hilbert space for $n>6$. }
  \label{fig:UDFidelity}
\end{figure}

We solve the master equation to evaluate the fidelity of the entangled qubit state at a finite temperature in the presence of damping of the CNT.  
 We choose the qubit state as $\ket{\Psi}\approx(0.52+0.51i)\ket{\uparrow \uparrow}+(-0.02i)\ket{\uparrow \downarrow}+(-0.02i)\ket{\downarrow \uparrow}+(-0.48+0.50i)\ket{\downarrow \downarrow}$ which can be obtained  at time $t_{\rm ideal}=76.8 \ { \rm \mu s}$ with initial state $\ket{\uparrow\uparrow}$ at zero temperature with the parameters in Fig. \ref{fig:concurrence}.  In  Fig. \ref{fig:Fidelity}, we plot the time evolution of the fidelity  with fixed $\Gamma=\omega_p /Q\approx 3\times10^5 \ {\rm s^{-1}}$ where $Q\approx30,000$ is reachable in experiment  \cite{Huettel2009, Steele2009}.  Because of the temperature $T=30\ {\rm mK}$, the fidelity of the case with $\Gamma =0$ is less than $1$.  From Fig. \ref{fig:Fidelity} (b), we can see the peak of the fidelity with  $\Gamma=0 $  is shifted from  $t_{\rm ideal}=76.8\ {\rm \mu s}$.  Although the iSWAP gate and the spin resonance mediated with virtual phonons are  not affected by the phonon numbers, the coefficient $\alpha$ of the rotation $\sigma_z$ at finite temperature is related to the phonons. The shift 
of the peak of the fidelity is larger at high temperature than at low temperature because the phonons follow Bose-Einstein statistics in the thermal bath (see Appendix  \ref{Appsec:fidelity}).  
 
We assume that the initial state is in the thermal equilibrium, so  damping does not change the phonon distribution. While a large damping rate slows down the transition process of the states with multiple phonons  whose spin-phonon coupling strength of the states with  phonon number $n$ is $ \sqrt{n} g$,  the fidelity  increases with an increasing damping rate.  
 In  Fig. \ref{fig:UDFidelity}, we plot the fidelity at  $t_{\rm ideal}=76.8\ {\rm \mu s}$  for $\ket{\Psi}$ with the initial state $\ket{\uparrow \uparrow}$ as a 
function of the damping rate.  While the damping rate increases, the fidelity surprisingly displays a minimum  when the damping rate approaches  $\Gamma  \approx 2\beta$ and  which we interpret as a stochastic resonance \cite{Huelga2007}. The damping $\Gamma$ of the phonon leads to transitions between configurations with the same qubit states and different phonon numbers.  The effective coupling $\beta$ is generated by the ac electric driving field with a large detuning as in Fig. \ref{fig:energylevel}.   From the effective coupling strength $\alpha_i$ in Eq.(\ref{eq:SchriefferInteraction}), one can see that the phonon number $n$  and therefore the effective coupling strength $\alpha_i$  fluctuate around their thermal average value with correlation time $1/\Gamma$.  These fluctuations do not significantly affect the coherent Rabi oscillation with Rabi frequency $\beta$, except when  their correlation time of the fluctuations is half of the period  of the driving field.  The stochastic resonance reaches its maximum at an optimal moderate value of the damping rate where $\Gamma   \approx  2\beta$\cite{Wellens2004}. On the other hand the ac electric driving field  is important for the electron spin resonance, but it does not have the effect of increasing phonon numbers for the large detuning $\Delta=\omega_{\rm p}-\omega$. With the stochastic resonance of the driving field and the damping, the transition between states with the same spin and different phonon numbers is enhanced,  which is detrimental for obtaining the electron spin resonance and the ideal maximally 
entangled states.  
Therefore, we find a minimal fidelity at the maximal stochastic resonance.

\section{Coupling of arbitrary QD pair in a QD array}
\label{sec:switch}

Universal quantum computation requires that arbitrary pairs of qubits can be coupled. We extend the case of two QDs to several QDs on the CNT, with one single electron trapped in each QD. To couple an arbitrary pair of QDs, the coupling between different pairs of QDs should be controllable. In  other words, we should be able to turn on and off the two qubit coupling between any arbitrary two qubits on the CNT. The coupling of two qubits is bridged by the spin-phonon interaction, hence it is possible to cut the coupling by breaking the spin-phonon coupling.  We show in the following that the interaction between any arbitrary two qubits can be switched off and on by controlling the spin-phonon coupling in each QD.

The spin-spin coupling in two QDs is induced by the inherent spin-phonon coupling in each QD.  
 As we have discussed in Sec.~\ref{sec:model},  under the precondition of a symmetric charge density function, the spin-phonon coupling is canceled if there is a symmetric distribution of the phonon waveform in the QD. By adiabatically changing the voltages which form the QDs, we can tune the location of the QDs to lie at the anti-nodes of the vibrational standing wave. When the phonon waveform is symmetric in the QD,  the spin-phonon coupling is eliminated. In other words, we can turn off the spin-phonon coupling by electrostatically shifting the electron wave function on the CNT. The left dot, for example, is between gates no. $1$ and no. $3$  in Fig. \ref{fig:switch} (b), therefore both distributions of the electron and the phonon waveform are symmetric   in the QDs and the coupling strength of the spin-phonon coupling is zero for each QD. To switch the interaction on, we can electrostatically shift the electron wave function to have an asymmetric phonon waveform in the QDs when  charge density 
function is symmetric. Therefore, arbitrary pairs of  QDs could be coupled from multiple QDs in CNT.

It is possible to produce the $X$, $Z$, and iSWAP gates, separately or together, by adjusting the strength and frequency of the driving field and the positions of the two QDs in CNT. A  series of one-qubit gates and iSWAP gates is sufficient for arbitrary quantum computation.

\section{Conclusions}

In summary, we have studied a nanomechanical system consisting of a suspended CNT where two separated single-electron spins in two QDs are coupled indirectly via the exchange of virtual phonons. The CNT is driven by an ac electric field in a parallel static magnetic field. The indirect coupling of the two spins is provided by the simultaneous coupling between the two spins and the vibrational mode of the CNT. We show that an iSWAP gate can be obtained by analyzing the effective Hamiltonian derived from the time dependent Schrieffer-Wolff transformation and the time evolution operator  when the driving electric field is off. Arbitrary single-qubit gates can be obtained in each QD by adjusting the strength and the frequency of the electric driving field. 
The iSWAP gate can be switched off when suppressing the spin-phonon coupling by electrostatically shifting the electron wave function on the CNT. 
  Combining the iSWAP gate and single-qubit gates in the double QDs in the CNT, a universal set of quantum gates can be built and 
  maximally entangled states of two spins can be generated in a single step by varying the frequency and the strength of the external electric driving field.  In this way, arbitrary pairs of distant spins in a QD array could be coupled.     The fidelity for obtaining a maximally entangled state at a fixed time  at finite temperature can be highly increased by increasing the damping rate of the CNT resonator.

\section*{Acknowledgements}

This work was supported by the DFG within the program SFB767. 
Heng Wang acknowledges a scholarship from the State Scholarship Fund of China.

\appendix

\section{Effective Hamiltonian from Schrieffer-Wolff transformation }
\label{Appsec:Schrieffer}

\begin{figure}[b]	
                   {\includegraphics[width=0.4\textwidth]{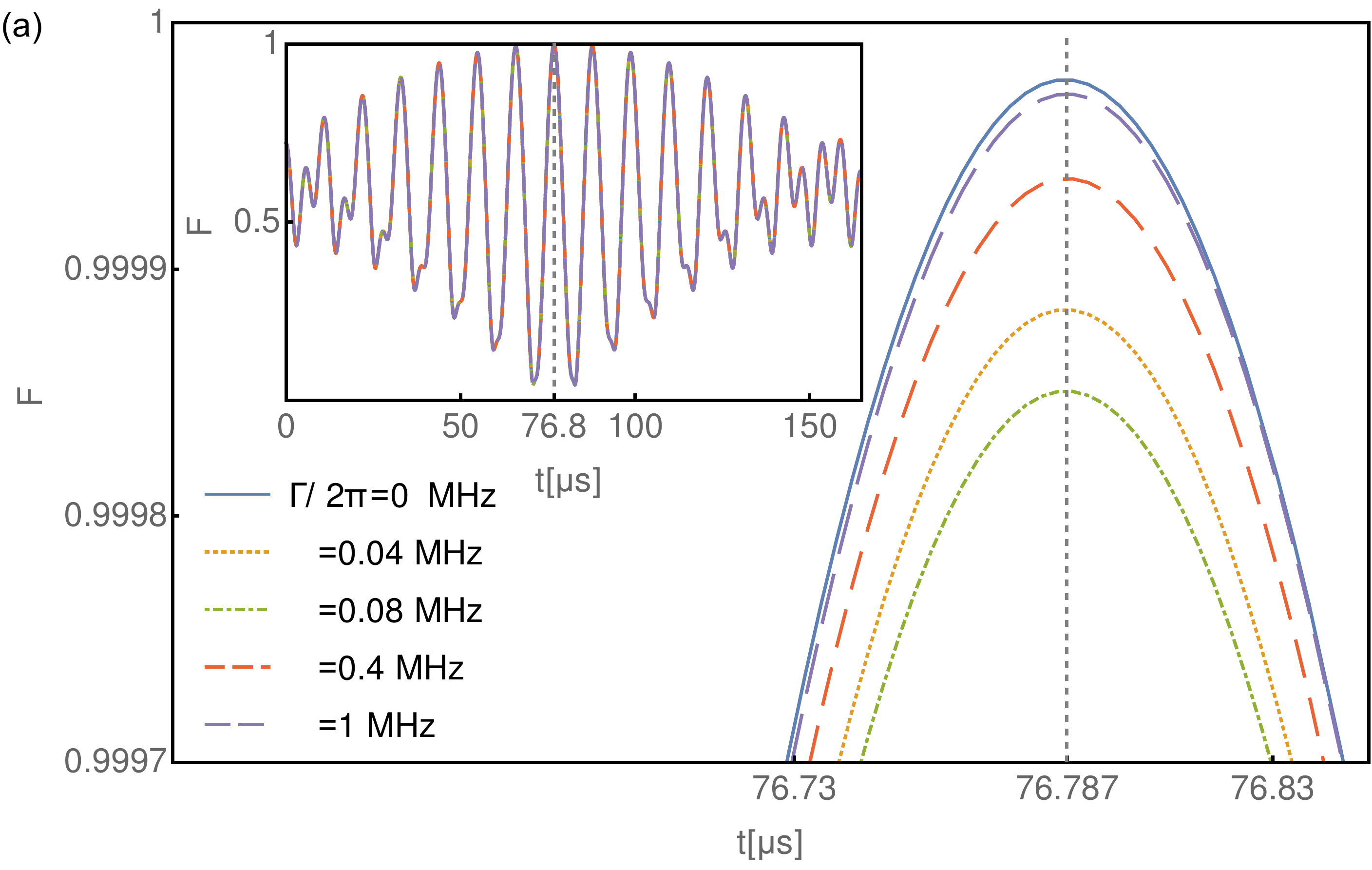}} 
                   {\includegraphics[width=0.4\textwidth]{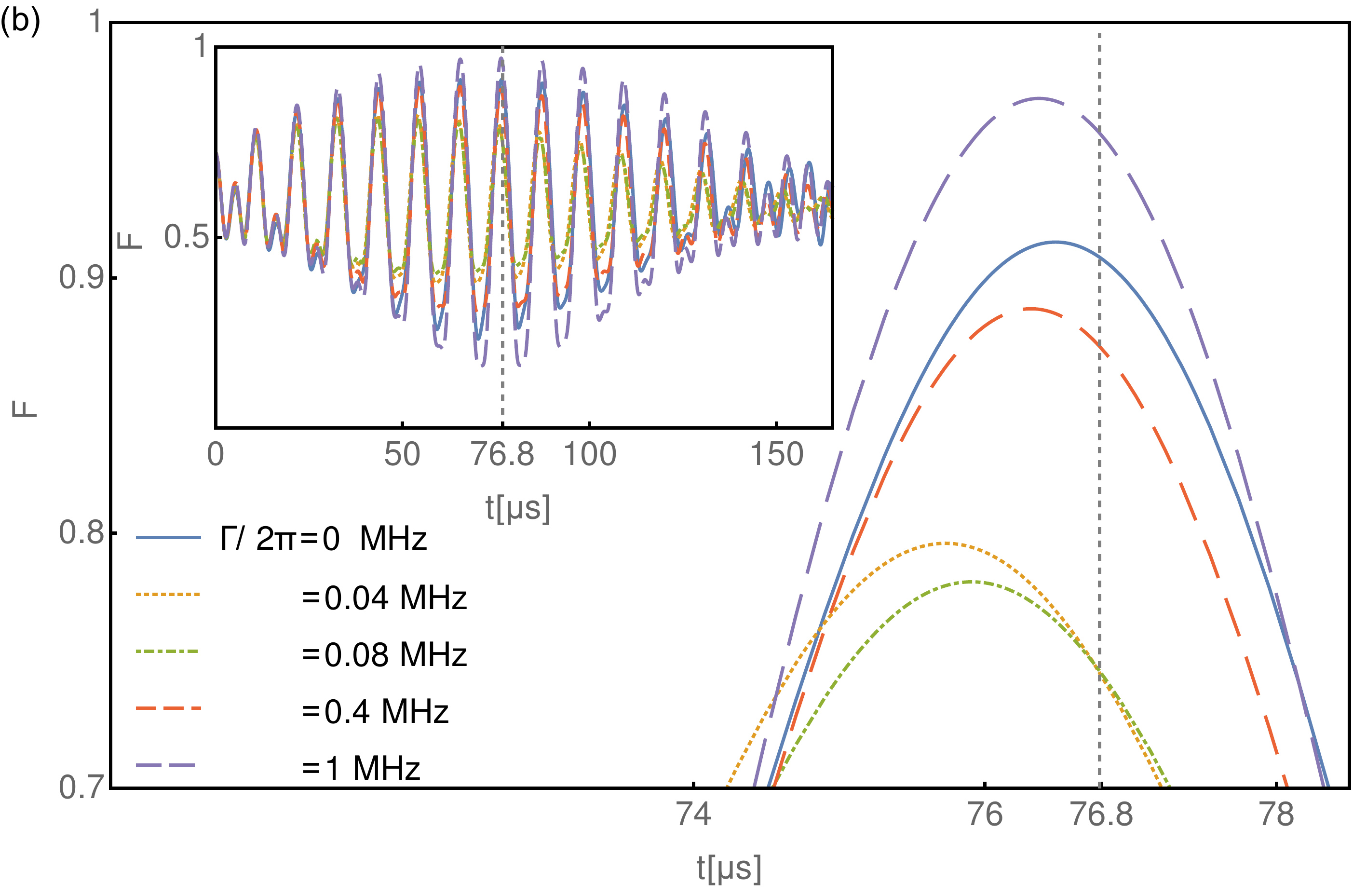}} 
\caption {Time evolution of the fidelity near $t_{\rm ideal}=76.8\ {\rm  \mu s}$ with driving strength $\lambda/(2\pi)=4.0 \ {\rm MHz}$ at finite temperature $T=10\ {\rm mK}$ in (a) and $T=100\ {\rm mK}$ in (b).   Inset: Fidelity over a larger time range; the circle denotes the area of the main plot. The other parameters are the same as in Fig. \ref{fig:concurrence}. }
  \label{fig:Fidelity100}
\end{figure}
In Appendix~\ref{Appsec:Schrieffer} we show how to obtain the effective Hamiltonian from the Schrieffer-Wolff transformation and obtain the form in the interaction picture.
We can obtain the time-dependent Schrieffer-Wolf transformation from the Schr\"odinger equation as 
\begin{equation}
H_{\rm eff}=UHU^\dag-iU(\partial_t U^\dag),
\label{eq:Sch}
\end{equation}
where $U(t)$ is  a unitary transformation. If one writes $U(t)=e^{S(t)}$, where $S(t)=-S(t)^\dagger \propto O(H_1)$, the transformed Hamiltonian at second order is 
\begin{equation}
H_{\rm eff}=H_0+O(H_1)+O(H_1^2).
\end{equation}
The first order terms $O(H_1)$ are eliminated to obtain a block-diagonal Hamiltonian that
\begin{equation}
H_{1}+[S(t),H_{0}]+i\dot{S}(t)=0,
\label{eq:firstorder}
\end{equation}
from which the expression of $S(t)$ can be obtained. Substitute the expression of $S(t)$ in Eq.~(\ref{eq:Sch}), we can obtain the transformed effective Hamiltonian $H_{\rm eff}$. 
 For simplicity, we assume $g_{A}=g_{B}=g$, $\omega_{\rm qi}=\omega_{\rm q}$, $\Delta=\omega_{\rm p}-\omega$ and $\Delta\gg g$. We obtain the effective Hamiltonian in the lowest subspace with $0$ phonon from the time-dependent Schrieffer-Wolff transformation in matrix form
\begin{widetext}
\begin{equation}
H_{\rm eff}=\left( 
\begin{array}{cccc}
\frac{\omega_{\rm q}(2g^2-\omega_{\rm p}^2+\omega_{\rm q}^2)}{-\omega_{\rm p}^2+\omega_{\rm q}^2} & -\frac{2\lambda g\omega_{\rm p}(\omega^2-2\omega_{\rm p}^2+\omega_{\rm q}^2)\cos{\omega t}}{(\omega^2-\omega_{\rm p}^2)(\omega_{\rm p}^2-\omega_{\rm q}^2)} & -\frac{2\lambda g\omega_{\rm p}(\omega^2-2\omega_{\rm p}^2+\omega_{\rm q}^2)\cos{\omega t}}{(\omega^2-\omega_{\rm p}^2)(\omega_{\rm p}^2-\omega_{\rm q}^2)}& -\frac{2 g^2 \omega_{\rm p}}{\omega_{\rm p}^2-\omega_{\rm q}^2}  \\
-\frac{2\lambda g\omega_{\rm p}(\omega^2-2\omega_{\rm p}^2+\omega_{\rm q}^2)\cos{\omega t}}{(\omega^2-\omega_{\rm p}^2)(\omega_{\rm p}^2-\omega_{\rm q}^2)}&0& -\frac{2 g^2 \omega_{\rm p}}{\omega_{\rm p}^2-\omega_{\rm q}^2}&-\frac{2\lambda g\omega_{\rm p}(\omega^2-2\omega_{\rm p}^2+\omega_{\rm q}^2)\cos{\omega t}}{(\omega^2-\omega_{\rm p}^2)(\omega_{\rm p}^2-\omega_{\rm q}^2)}\\
-\frac{2\lambda g\omega_{\rm p}(\omega^2-2\omega_{\rm p}^2+\omega_{\rm q}^2)\cos{\omega t}}{(\omega^2-\omega_{\rm p}^2)(\omega_{\rm p}^2-\omega_{\rm q}^2)}&-\frac{2 g^2 \omega_{\rm p}}{\omega_{\rm p}^2-\omega_{\rm q}^2}& 0&-\frac{2\lambda g\omega_{\rm p}(\omega^2-2\omega_{\rm p}^2+\omega_{\rm q}^2)\cos{\omega t}}{(\omega^2-\omega_{\rm p}^2)(\omega_{\rm p}^2-\omega_{\rm q}^2)}\\
-\frac{2 g^2 \omega_{\rm p}}{\omega_{\rm p}^2-\omega_{\rm q}^2} &-\frac{2\lambda g\omega_{\rm p}(\omega^2-2\omega_{\rm p}^2+\omega_{\rm q}^2)\cos{\omega t}}{(\omega^2-\omega_{\rm p}^2)(\omega_{\rm p}^2-\omega_{\rm q}^2)}& -\frac{2\lambda g\omega_{\rm p}(\omega^2-2\omega_{\rm p}^2+\omega_{\rm q}^2)\cos{\omega t}}{(\omega^2-\omega_{\rm p}^2)(\omega_{\rm p}^2-\omega_{\rm q}^2)}&-\frac{\omega_{\rm q}(2g^2-\omega_{\rm p}^2+\omega_{\rm q}^2)}{-\omega_{\rm p}^2+\omega_{\rm q}^2}
\end{array} \right).
\label{eq:SchriefferEffMatrix}
\end{equation}

We transform Eq.~(\ref{eq:SchriefferEffMatrix}) into the interaction picture with respect to  $H_0$. 
 The fast oscillating terms with $e^{\pm i(\omega +\omega_{\rm q})t} $ and $e^{\pm 2i\omega_{\rm q} t} $ can be dropped in the rotating-wave approximation. We find
 
\begin{equation}
\begin{split}
H_{\rm eff}^{I}/\hbar=&(\frac{\omega_{\rm q} g^2}{-\omega^2_{\rm p}+\omega_{\rm q}^2})(\sigma_{Az}
+\sigma_{Bz})\\
&-\frac{\lambda g\omega_{\rm p}(\omega^2-2\omega_{\rm p}^2+\omega_{\rm q}^2)}{(\omega^2-\omega_{\rm p}^2)(\omega_{\rm p}^2-\omega_{\rm q}^2)}(e^{i\omega t}+e^{-i\omega t})(\sigma_{A+}e^{i\omega_{\rm q} t}+\sigma_{B+}e^{i\omega_{\rm q} t}+\sigma_{A-}e^{-i\omega_{\rm q} t}+\sigma_{B-}e^{-i\omega_{\rm q} t})
\\&-\frac{2 g^2 \omega_{\rm p}}{\omega_{\rm p}^2-\omega_{\rm q}^2} (\sigma_{A+}\sigma_{B-}+\sigma_{A-}\sigma_{B+}+\sigma_{A+}\sigma_{B+}e^{2i\omega_{\rm q} t}+\sigma_{A-}\sigma_{B-}e^{-2i\omega_{\rm q}t}).
\label{eq:SchriefferInteraction1}
\end{split}
\end{equation}
\end{widetext}
We obtain the effective   Hamiltonian   in the rotating frame with $ U=e^{i\omega t}$:
\begin{equation}
\begin{split}
H_{\rm eff}^{I}/\hbar=&-\frac{(\omega_{\rm p}-\Delta) g^2}{(2\omega_{\rm p}-\Delta)\Delta}(\sigma_{Az}
+\sigma_{Bz})\\
&-\frac{\lambda g 2 \omega_{\rm p}}{(2\omega_{\rm p}-\Delta)\Delta}(\sigma_{Ax}+\sigma_{Bx})
\\&-\frac{2 g^2 \omega_{\rm p}}{(2\omega_{\rm p}-\Delta)\Delta}( \sigma_{A+}\sigma_{B-}+ \sigma_{A-}\sigma_{B+}).
\end{split}
\label{eq:SchriefferInteractionII}
\end{equation}

\section{Fidelity with $T=10\ {\rm mK} $ and $T=100\ {\rm mK} $}
\label{Appsec:fidelity}
 
When we consider the case with the phonon bath and the damping effect, the detunings between the peak of the fidelity without damping and the ideal time point $T_{ideal}= 76.8\ {\rm \mu s}$ increase while the temperature increases. We can  compare the two cases  at $T=10 \ {\rm mK}$ and $T=100  \ {\rm mK}$ in Fig.~\ref{fig:Fidelity100}. The shifting is obvious at $T=100 \  {\rm mK}$ but very small at $T=10 \ {\rm mK}$. This is because the phonons follow the Bose-Einstein distribution in the thermal bath and the coefficient  $\alpha_i$ of the rotation  $\sigma_z$ in the effective Hamiltonian in Eq.~(\ref{eq:SchriefferInteraction}) is related to the phonons.

\bibliographystyle{apsrev}
\bibliography{citeinteraction}

\end{document}